# High frequency GaAs nano-optomechanical disk resonator

Lu Ding[1], Christophe Baker[1], Pascale Senellart[2], Aristide Lemaitre[2], Sara Ducci[1], Giuseppe Leo[1], Ivan Favero[1]*

[1]Laboratoire Matériaux et Phénomènes Quantiques, Université Paris Diderot, CNRS, UMR 7162, 10 rue Alice Domon et Léonie Duquet, 75013 Paris, France
[2]Laboratoire de Photonique et Nanostructures, CNRS, Route de Nozay, 91460 Marcoussis, France
* ivan.favero@univ-paris-diderot.fr

**Abstract:** Optomechanical coupling between a mechanical oscillator and light trapped in a cavity increases when the coupling takes place in a reduced volume. Here we demonstrate a GaAs semiconductor optomechanical disk system where both optical and mechanical energy can be confined in a sub-micron scale interaction volume. We observe giant optomechanical coupling rate up to 100 GHz/nm involving picogram mass mechanical modes with frequency between 100 MHz and 1 GHz. The mechanical modes are singled-out measuring their dispersion as a function of disk geometry. Their Brownian motion is optically resolved with a sensitivity of $10^{-17}$m/√Hz at room temperature and pressure, approaching the quantum limit imprecision.

Optomechanical systems generally consist of a mesoscopic mechanical oscillator interacting with light trapped in a cavity [1-3]. These systems have attracted a growing interest since first experimental evidences that cavity light can be used to optically self-cool the oscillator towards its quantum regime [4-9]. They are now studied in an increasing number of geometries and compositions, with the common purpose of coupling photons and phonons in a controlled way. Beyond the mere goal of reaching the quantum ground state of a mechanical oscillator, today concepts developed in optomechanics find applications in very different fields such as cold atoms physics [10-11], mechanical sensing [12] or Josephson circuitry [13]. High-frequency nanomechanical oscillators are generally welcome in these applications, to ease the access to the quantum regime or to develop high-speed sensing systems. However since their sub-wavelength size generally imply a weak interaction with light, these oscillators need to be inserted in a cavity to enhance the optical/mechanical interaction [14]. The typical optomechanical coupling obtained using this approach is of 10 MHz/nm for visible photons [15-16] or 10 kHz/nm in the microwave range [17]. A coupling enhancement can be obtained by further confining mechanical and optical modes in a small interaction volume, as recently achieved in nano-patterned photonic crystals [18]. However these structures are complex to design and fabricate, and being based on silicon technology, they do not allow the insertion of an optically active medium. This precludes exploring novel situations where a (quantum) mechanical oscillator would be coupled to a (quantum) photon emitter embedded in the host material. In this paper we present a gallium arsenide (GaAs) nano-optomechanical disk resonator, a system at the crossroads with III-V semiconductor nano-photonics. This resonator combines the assets of both nano-scale mechanical systems (high frequency and low mass in the pg range) and semiconductor optical micro-cavities, with optical quality factor above $10^5$. The high refractive index of GaAs enables storing light in a sub-micron mode-volume whispering gallery mode of the disk, where it couples to high frequency (up to the GHz) vibrational modes of the structure. Thanks to the miniature optical/mechanical interaction volume, the coupling reaches 100 GHz/nm, providing nearly quantum-limited optical detection of vibrations. Besides offering high frequency mechanical oscillators with an ultra-sensitive optical read-out, GaAs nano-optomechanical resonators are naturally suited for integration in geometry of arrays on a chip and for interfacing with single emitters in form of InGaAs quantum dots.

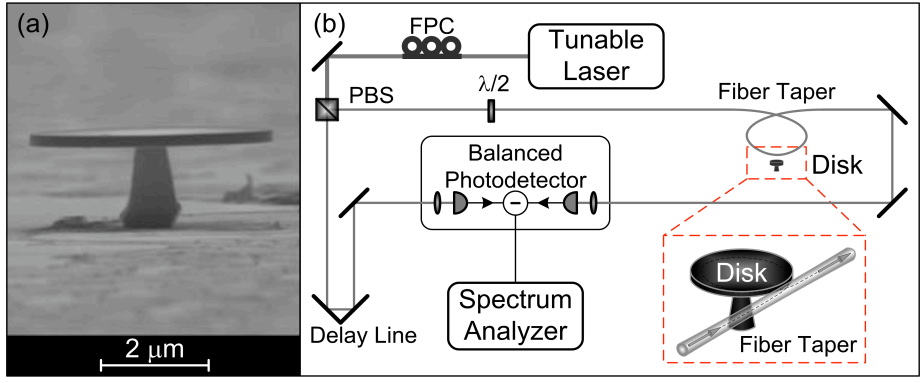

Fig. 1. Optomechanical study of a GaAs disk. a) Scanning Electron Microscope (SEM) view of a GaAs disk (4.5 µm diameter and 200 nm thickness) suspended on an AlGaAs pedestal. b) Schematics of the near-field optomechanical spectroscopy experiment. FPC stands for Fiber Polarization Control, PBS for Polarization Beam Splitter.

A whispering gallery disk structure exhibits the salient feature of being both an optical cavity and a mechanical oscillator. The radial breathing of the structure couples naturally to photons stored in the gallery mode, by modulating the effective optical length of the cavity. The related optomechanical coupling factor $g=(d\omega/d\alpha)$ accounts for the differential dependence of the cavity angular eigenfrequency $\omega$ on the mechanical displacement $\alpha$, and enters the quantum description of the system coupled through radiation pressure $H_c= \hbar g(\hat{a}^+\hat{a})\alpha$, where $\hat{a}$ ($\hat{a}^+$) is the optical annihilation (creation) operator and $\alpha$ the mechanical degree of freedom. A large value of g is required for efficiently turning mechanical information into optical information and vice-versa, and is hence beneficial to most applications in optomechanics experiments: sensitive optical measurement of the mechanical oscillator displacement, stiffening or softening of the oscillator spring through the optical spring effect, and optical dynamical back-action on the oscillator, responsible for its self-cooling or self-oscillation. As we will see below, a miniature whispering gallery structure can sustain very large value of g.

The optical whispering gallery mode (WGM) problem can be treated solving the Helmholtz equation in a cylinder of height h and radius R [19]. Provided that h is sufficiently larger than $\lambda/n$ (thick disk limit) the effective index method can be employed to separate vertical (along cylinder axis) and horizontal dependence of the electromagnetic field F (F=E or H). Using the rotational invariance, F is decomposed $F=\Psi(\rho)\Theta(\theta)G(z)$ with $\Theta(\theta)=e^{im\theta}$. E.g. for the TM modes of azimuthal number m, the dispersion relation imposed by continuity of tangential E and H at the disk interface reads $n_\zeta(\dot{J}_m(kn_\zeta R) / J_m(kn_\zeta R)) - \dot{H}_m^{(2)}(kn_\zeta R) / H_m^{(2)}(kn_\zeta R) = 0$ where $n_\zeta$ is the effective index of the slab of thickness h, $J_m$ is the first-kind Bessel function of order m, $H^{(2)}_m$ is the second-kind Hankel function of order m, and $k=2\pi/\lambda$. This dispersion relation only depends on kR: as a consequence, if we have a solution $k_0$ to this equation for a radius $R_0$, the solution k for radius R is given by $k=k_0 R_0/R$. Thus in the limit of a thick disk, $d\omega/dR=-\omega/R$ is the exact optomechanical coupling factor g for a pure radial displacement dR. Thanks to a large refractive index, GaAs whispering gallery structures with radius as small as 1 µm can sustain high quality factor (Q) optical WGMs in the near infrared [20-22]. According to the formula $g=-\omega/R$, g values are expected to rise up to the THz/nm range on such disks.

Our GaAs disks are fabricated from an epitaxial wafer using e-beam lithography and wet etching [21]. The typical disk size is 5 µm in diameter and 200 nm in thickness, as can be seen in Fig. 1a. Near-field optical experiments on a single disk are performed using an optical fiber taper evanescent coupling technique (Fig. 1b). Infrared laser light is coupled in an optical fiber whose tapered part is evanescently coupled to the disk. Photons are collected on a photodetector at the output of the fiber. First, a single-mode telecom optical fiber with 10 (125) µm core (cladding) diameter is adiabatically stretched down to a ~ 800 nm diameter taper using a microheater [23], so that its evanescent field can access the surroundings. The fiber taper is subsequently curled into a micro-loop structure with typical diameter 70 µm. This micro-loop improves spatial selectivity in the experiments by acting as a near-field optical "point-probe" [23]. Using XYZ piezo translation stages, the looped fiber taper is positioned within the disk near-field to allow evanescent coupling with an adjustable gap distance between disk and taper (see inset of Fig. 1b and details in [24]). We use a tunable external cavity diode laser ($\lambda$=1500–1600 nm) with linewidth smaller than 1 MHz). In our set-up, the tapered fiber is mounted in one arm of a balanced detection path, with a half-wave plate used to select the injected linear polarization. The second arm consists of a free beam path. The two arms are recombined on a high-speed balanced photodetector and the difference signal analysed by a spectrum analyser. When tuning the laser wavelength to a flank of the disk optical resonance (inset of Fig. 2a), the disk vibration modulates the optical transmission T of the tapered fiber and the set-up performs optical vibrational spectroscopy of the disk. The sensitivity of the measurement relies on the efficient transduction of the mechanical displacement into a variation of the transmission T. On a optical resonance flank and in the limit of a small displacement $\Delta\alpha$, we have T($\Delta\alpha$)-T(0)=(dT/d$\omega$)×(d$\omega$/d$\alpha$)=$\Delta$T×Q×g where $\Delta$T=1-$T_{on}$ is the contrast of the optical WGM resonance, $T_{on}$ the on-resonance transmission and Q the loaded optical quality factor, which is typically in the range of $10^5$ in our experiments. $\Delta$T and Q being measured quantities, the calibration of a vibrational measurement directly gives the optomechanical coupling factor g.

Figure 2 shows vibrational optical noise spectra obtained on selected disks using the above-described technique. The measurements are performed at room temperature and pressure. Several mechanical resonances are observed with amplitude up to 20dB over the noise floor. The balanced photodetection enables removing the contribution of laser excess noise in the measurement. A first striking feature in the spectra is the high frequency of the observed mechanical resonances, which span from 100 MHz to 1 GHz. This results from the small size of the disk mechanical oscillators, which also induces small motional masses (see also Table 1). The mechanical Q factors span between a few tens and a few $10^3$. Fig. 2c shows for example a mechanical resonance at 858.9 MHz with Q=862. This corresponds to a frequency-Q product of $0.7\times10^{12}$, approaching best reported values in the $10^{13}$ range [25] and obtained here in ambient conditions. The observed mechanical Q results of course from added contributions of air damping, clamping losses, thermoelastic damping, that could all be reduced by an appropriate design of the disk clamping and by operating the system under vacuum at low temperature. Moreover we stress that we obtain crystalline GaAs material of extreme purity by Molecular Beam Epitaxy (MBE) techniques. Our crystalline mechanical oscillators should hence be free of bulk loss mechanisms affecting amorphous glasses like Silica and Silicon Nitride at low temperature [26]. A GaAs micromechanical oscillator with Q factor above $10^5$ has already been reported at low temperature [27]. Mechanical dissipation in GaAs nano-optomechanical disk resonators will be investigated in future work.

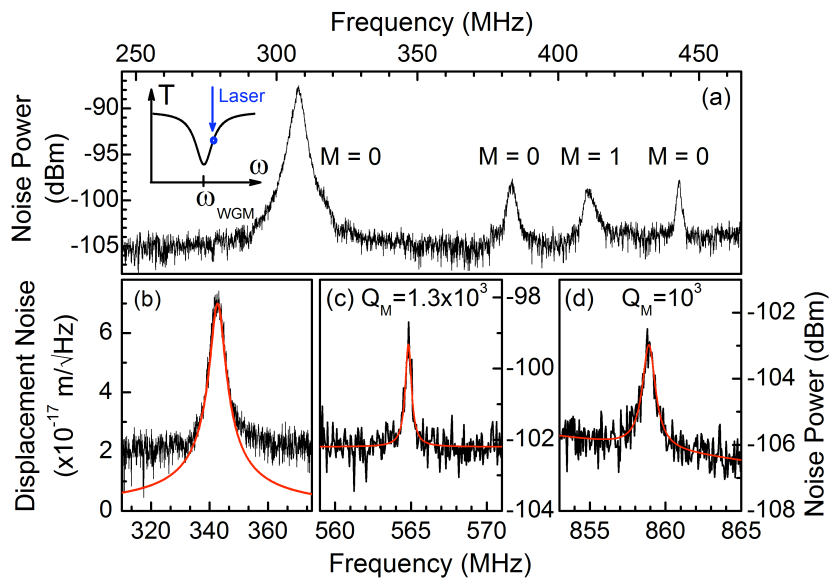

Fig. 2. Selected vibrational spectra of GaAs nano-optomechanical disks. a) Optically measured motional noise spectrum of a disk of radius 5.6 µm and thickness 200 nm. b) Calibrated displacement noise resonance of a disk of radius 4.5 µm, showing shot-noise limited sensitivity of $2\times10^{-17}$m/√Hz. c) High frequency mechanical resonances of a disk of radius 3.6 µm in the 500MHz-1GHz band.

Fig. 2b displays a measurement of the disk Brownian motion at 300K, showing a sensitivity of $2\times10^{-17}$m/√Hz. The calibration is obtained here using equipartition of energy for the Brownian motion amplitude but it can also be obtained from an independent estimation of g, as detailed below.

The formula g=-ω/R is expected to be no longer valid if the mechanical mode has a non-radial component (out-of-plane motion) or in case the effective index approach does not hold anymore. The latter occurs if the disk is too thin (<λ/n) or if it is not a perfect cylinder. With a 200 nm thickness, we expect our GaAs disk resonators to deviate importantly from g=-ω/R. The general problem of how an optical resonator eigenfrequency ω depends on the deformation α of the resonator can be solved by a perturbative treatment of Maxwell's equations [28]. A convenient approach is to represent the resonator deformation by the displacement α of a chosen point of the resonator having maximum displacement amplitude. If the normalized displacement profile is known, α suffices to represent the complete resonator deformation field. With this approach g is expressed as the integral of an optomechanical function over the rigid resonator boundaries:

$$g = \frac{\omega_0}{4} \times \int (\vec{q}\cdot\vec{n})\left[\Delta\varepsilon\left|\vec{e}_\parallel\right|^2 - \Delta(\varepsilon^{-1})\left|\vec{d}_\perp\right|^2\right]dA \quad (1)$$

In this expression $\omega_0$ is the rigid resonator eigenfrequency, **q** is the normalized displacement profile vector such that max|**q**(r)|=1, n the normal unit vector on the boundary, Δε=ε-1 with ε the dielectric constant of the resonator material, $\Delta(\varepsilon^{-1})=\varepsilon^{-1}$-1, **e** is the electric complex field vector normalized such that ½ ∫ ε|**e**|²dV=1 and **d**=ε**e**. To validate Eq. 1 in a simple case, we computed numerically whispering gallery modes profiles of a thick (≥λ/n) GaAs disk surrounded by air using the Finite Element Method Comsol software [29]. We adopted a pure radial displacement profile, free of any out of plane motion, and calculated g using Eq. 1. The results agree with the thick disk limit g=-ω/R within typically 1%. We also checked the validity of Eq. 1 in the case of a standard Fabry-Pérot and note that it was successfully used to model properties of optomechanical crystals structures [18]. Importantly Eq. 1 shows that g depends both on the electromagnetic and mechanical mode under consideration. It is incorrect to assume that a single optomechanical coupling g holds true for all mechanical modes of a given resonator. Such simplification can lead to important errors in the calibration of optomechanical measurements. In the case of interest here, both electromagnetic and mechanical modes of the GaAs disk are identified as explained now.

The WGM identification is carried out performing broadband optical spectroscopy of the disk to reveal a sequence of WGM resonances of varying linewidth and wavelength spacing. This sequence is compared to that obtained from numerical simulations, in which each WGM of radial p and azimuthal m numbers is assigned a resonance wavelength and a radiative linewidth. The overlap of the two sequences (measured and simulated) is generally excellent and imposes a unique label (p,m) to each optical resonance observed in the spectrum [24]. For example, the WGM used in the measurement of Fig. 2b is found to be the (p=3, m=33) WGM of the disk.

The identification of mechanical modes appearing in the measurement is more involved. Indeed, their frequencies depend on the exact shape and dimensions of the disk and pedestal, on the possible presence of residual stress in the material after etching, and on the exact composition of the AlGaAs pedestal. To avoid misidentification of the modes induced by an inexact knowledge of some of these parameters, we study disks of different size but on the same sample, hence made out of the same material and resulting from the same fabrication process. This allows studying the dispersion line of each identified mechanical mode as a function of the disk size. We start with a thorough SEM inspection of the disks, with special attention to the disk diameter and geometry of the supporting pedestal. The disk thickness and pedestal height are controlled during epitaxial growth. The other geometrical parameters are measured with a precision of ± 50 nm and serve as input for numerical simulations, together with bulk elastic parameters for GaAs (Young Modulus E=85.9 GPa, Poisson Ratio σ=0.31 and Density 5.316 g.cm$^{-3}$) and $Al_{0.8}Ga_{0.2}As$ (E=83.86 GPa, σ=0.39 and Density 4.072 g.cm$^{-3}$). These simulations provide us for each disk with a series of mechanical modes that can be named using radial and azimuthal numbers P and M. The displacement profile of a given mode can be factorized in the form $q(\rho,\theta,z)=\cos(M\theta) \times f_p(\rho,z)$ as a result of the rotational symmetry of the disk.

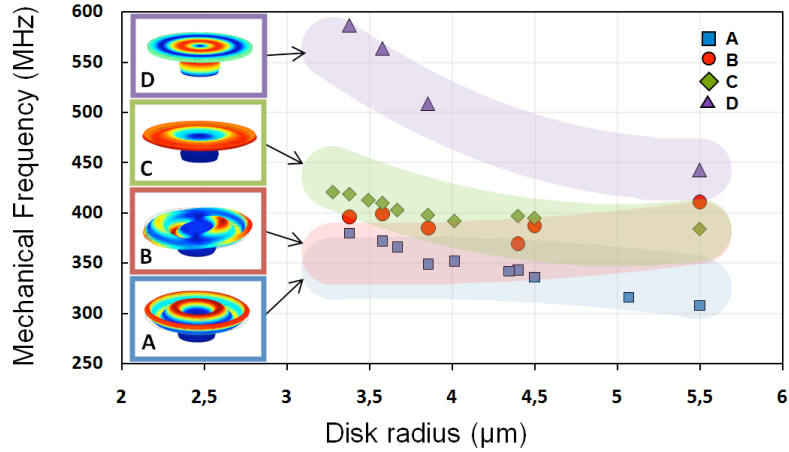

Fig. 3. Dispersion lines of GaAs disk mechanical modes measured vs. disk radius. Thickness is 200nm. Data points were acquired on different disks of the same sample. Transparent "guide to the eyes" bands correspond to numerical simulations with an error bar represented by their thickness.

Fig. 3 shows measurements performed on a series of disks of radius varying between 3 and 6 µm. The shadowed guides to the eyes on the figure are obtained from simulations and displayed with a thickness of ±30 MHz corresponding to our uncertainty in the measurement of the disk geometrical parameters. The agreement obtained in Fig. 3 allows identifying the observed mechanical resonances in the spectrum as having M=0 and M=1 azimuthal numbers. This identification is further confirmed by evaluating the related optomechanical coupling. Indeed as a consequence of the disk rotational symmetry, the integral of Eq.1 contains an azimuthal pre-factor of the form:

$$\int_{\theta=0}^{2\pi} \cos(M\theta)\cos^2(m\theta)d\theta = \frac{1}{2}\delta_{M,0} + \frac{1}{4}\left[\delta_{M+2m,0} + \delta_{M-2m,0}\right] \quad (2)$$

where m (M) is the azimuthal number of the optical (mechanical) mode under consideration. From Eq.2 it is obvious that the largest optomechanical coupling g is obtained for the mechanical azimuthal number M=0. This is confirmed by our measurement where dominating resonances in the spectra always correspond to M=0 modes.

In case of interest we have m>1 and $\delta_{M+2m,0}$=0. Apart from the case M=0, the only other case of non-zero optomechanical coupling is thus expected for M=2m. In WGMs investigated here, m goes from 30 to 50 and mechanical modes with to M=2m are hence of very high order. Because their displacement field is of small amplitude, they are practically not visible in the spectra with our current sensitivity. In contrast, we observe mechanical modes with small azimuthal numbers (M=1,2...), even though their weight in the spectrum remains small compared to M=0 modes. The presence of M=1 modes in the optical measurement shows that the rotational symmetry of the disk and the selection rules imposed by Eq.2 are only approximate. The disk symmetry can be broken by an asymmetry of the pedestal, an ellipticity of the disk or even by the residual roughness of the disk sidewall. Sidewall roughness is for example responsible of the lifting of degeneracy of clockwise and counter-clockwise WGMs [24,30]. In our case, we have observed that some disk pedestals end-up being slightly rectangular after wet etching, which we anticipate to be the dominating source of asymmetry.

Once both optical and mechanical modes are singled out in the measurement, the optomechanical coupling g is computed for each optical/mechanical combination using Eq. 1. The mechanical resonance shown in Fig. 2b corresponds for example to an M=0 mode that we have labeled A in Fig. 3. The optical resonance in Fig. 2b has been identified to be the WGM (p=3, m=33). A numerically "simulated" value $g^s$ of the coupling between these two modes is computed and found to be $g^s$=61 GHz/nm. But the effective mass of the mechanical mode also leads the Brownian motion amplitude by the equipartition theorem $m_{eff}\omega^2<\Delta\alpha^2>=k_BT$. Our measurements are performed at 300 K and the magnitude of the thermo-optic shift of the WGM resonance indicates that heating of the disk induced by the laser remains lower than 5 K, which induces negligible modification of the thermal motion amplitude [31]. The Brownian motion can then also be used to calibrate the measurement and to extract in a second independent manner an "experimental" value $g^e$ of the optomechanical coupling. $g^e$ and $g^s$ are generally in reasonable agreement.

Table 1. Selected optical/mechanical resonances on 3 GaAs disks. Experimental (superscript e) and simulated (superscript s) parameters are given.

| $R_{disk}$ (μm) | $\lambda_O$ (nm) (p,m) | Optical Q | $f_M^e$ (MHz) | $f_M^s$ (MHz) | $m_{eff}^s$ (pg) | $g_{om}^e$ (GHz/nm) | $g_{om}^s$ (GHz/nm) | Mechanical Q |
|---|---|---|---|---|---|---|---|---|
| 4.5 ± 0.1 | 1551.7 (3,33) | $1.0 \times 10^5$ | 343 | 339 | 21.2 | 92 | 61 | 66 |
|  |  |  | 397 | 380 | 21.2 | 85 | 76 | 38 |
| 4.6 ± 0.1 | 1535.2 (3,34) | $9.5 \times 10^4$ | 336 | 337 | 22.1 | 124 | 65 | 120 |
|  |  |  | 395 | 380 | 20.9 | 77 | 72 | 66 |
| 5.6 ± 0.1 | 1513.9 (3,45) | $5.9 \times 10^4$ | 308 | 326 | 36 | 172 | 78 | 73 |
|  |  |  | 383.5 | 373 | 22.4 | 53 | 35 | 114 |

Table 1 contains the parameters relating to optical/mechanical resonances observed on 3 distinct disks. The effective masses are typically around 10 pg, with optomechanical coupling spanning from 50 to 200 GHz/nm. Combined with an optical Q factor in the $10^5$ range, this leads to a typical displacement sensitivity of $10^{-17}$m/√Hz. We stress that this is only a factor 100 above the Standard Quantum Limit imprecision and that this is obtained here at room temperature and ambient pressure, which is an important advantage for future sensing applications. This sensitivity compares favourably with other systems considering the nanoscale dimensions of the GaAs mechanical system under consideration and the high mechanical frequencies at play. Previous displacement sensitivities on MHz nanomechanical oscillators are of $10^{-16}$m/√Hz obtained with a single electron transistor at mK temperature [32], $10^{-15}$m/√Hz using a RF strip-line cavity at mK temperature [17] and $10^{-15}$m/√Hz reachable using a high finesse fiber-cavity or a silica toroid as a mechanical/optical transducer [15-16]. A sensitivity in the $10^{-17}$m/√Hz range was very recently reported in optomechanical crystals structures [18].

This sensitivity makes semiconductor nano-optomechanical disk resonators promising candidates to study fundamental quantum limits of displacement measurement. For example, a GHz mechanical mode of a GaAs disk would have a mean thermal occupation number of only 35 at 1K, on par with values reported recently using a combination of cryogenics and self-cooling techniques on oscillators of lower frequency [33-35]. Cavity self-cooling of such disk oscillator by a factor as low as 35 would allow it to enter the quantum regime, with the peculiar advantage that it would be available on a integrable optical platform. Besides the envisioned coupling of this quantum mechanical oscillator to an artificial atom (InAs quantum dot) for coherent control experiments, the GaAs-based optomechanical system presented here lends itself to the design of optical and piezo-electrical properties through doping and quantum engineering of an active medium. These original features should allow exploring novel architectures, at the frontier of III-V nanophotonics and optomechanics in the quantum regime.

This work has been supported by the Region Ile-de-France in the framework of C'Nano IdF.